\documentclass[aps,prd,onecolumn,groupedaddress,showpacs,nofootinbib,amssymb]{revtex4}
\usepackage[dvips]{graphicx}
\usepackage{amssymb}
\usepackage{amsmath}
\usepackage{graphicx,,color}
\usepackage{amsfonts}
\usepackage{bm}
\usepackage{cancel}
\usepackage{comment}
\usepackage[dvipsnames]{xcolor}
\usepackage{makecell}
\usepackage{amsmath}
\usepackage{amsfonts}
\usepackage{physics}

\allowdisplaybreaks[4]

\begin{document}

\tolerance=5000

\title{Recombination Thickness as an Uncertainty in Inflationary Observables}
\author{V.K. Oikonomou$^{1,2}$}\email{voikonomou@gapps.auth.gr;v.k.oikonomou1979@gmail.com}
\affiliation{$^{1)}$Department of Physics, Aristotle University of
Thessaloniki, Thessaloniki 54124, Greece\\
$^{2)}$ Center for Theoretical Physics, Khazar University, 41
Mehseti Str., Baku, AZ-1096, Azerbaijan}

\begin{abstract}
Standard CMB analysis assumes a direct deterministic mapping
between the multipole probed by the CMB $\ell$ and the primordial
wavenumber $k$. Since the recombination era has a finite duration,
this mapping is probabilistic by construction. We elevate the
power spectrum of the primordial perturbations to a probability
distribution caused by  the finite duration of the recombination
era. We show that a finite recombination width introduces a
Gaussian smoothing scale in $\ln k$ with $\sigma_{\ln k} \sim
\sigma_\eta / D_*$, leading to a probabilistic mapping from
multipoles to inflationary e-folds. This effect is zero in
standard power-law inflationary scenarios, but it may become
relevant for scenarios with exotic oscillating features of the
primordial power spectrum, which will be probed by the future CMB
experiments. The observed effective power spectrum is the true
primordial spectrum blurred by the uncertainty in scale
reconstruction, which is mathematically identical to a Bayesian
marginalization over a latent variable, and thus there is a
propagation of the measurement error in the independent variable,
which is another more formal way to view the smoothing effect. Our
results indicate that the smoothing has quantifiable effects on
the spectral index and its running, but more importantly the
difference between the TT and EE inferred spectral indices,
$n_s^{TT}-n_s^{EE}$, is non-trivial, in contrast to standard
inflation without smoothing, and might become observable by future
cosmic microwave background experiments. Any tension in
$n_s^{TT}-n_s^{EE}$ could indicate oscillations in the primordial
spectrum and the effects of the power spectrum smoothing. Finally,
a minimal Fisher matrix analysis is performed to investigate the
observability prospects of the smoothing effect.
\end{abstract}

\maketitle

\section{Introduction}

In the next decade, the primordial era of our Universe, which is
the most fundamental era, will be severely scrutinized. Inflation
\cite{inflation1,inflation2,inflation3,inflation4}, is the
prominent scenario for this mysterious era of our Universe, and
the hope is that the cosmic microwave background (CMB) radiation
experiments will reveal hints for the theory that drives the
inflationary regime, or even if quantum gravity effects are
involved in this classical epoch. The CMB oriented experiments
like the Simons observatory \cite{SimonsObservatory:2019qwx} and
the LiteBird \cite{LiteBIRD:2022cnt}, are expected to further
constrain the inflationary era, or even detect direct probes of
the inflationary regime, like for example detecting the B
polarization modes of the CMB. This detection would be a smoking
gun signal for the inflationary regime. On the other hand,
imprints of the inflationary era could be revealed by the
existence of a stochastic gravitational wave background of
cosmological origin, and this will be sought by the future
gravitational wave experiments
\cite{Hild:2010id,Baker:2019nia,Smith:2019wny,Crowder:2005nr,Smith:2016jqs,Seto:2001qf,Kawamura:2020pcg,Bull:2018lat,LISACosmologyWorkingGroup:2022jok}.
The existence of such a stochastic background is now a scientific
fact, as it was confirmed by the Pulsar Timing Array experiments
like NANOGrav \cite{NANOGrav:2023gor} but the signal itself cannot
be explained solely by inflation
\cite{Vagnozzi:2023lwo,Oikonomou:2023qfz}.

The primordial spectrum might not be a pure power-law but can be
oscillating
\cite{Zeng:2018ufm,Domenech:2019cyh,Danielsson:2002kx,Jackson:2013vka,Kempf:2000ac,Easther:2001fz,Martin:2003kp,Chen:2008wn,Flauger:2009ab,Covi:2006ci,Hamann:2007pa,
Hazra:2010ve,Shafieloo:2003gf,Achucarro:2013cva,Hazra:2014jwa,Chen:2014cwa,Nicholson:2009pi,Hunt:2015iua,Braglia:2021ckn,Braglia:2021sun,Antony:2021bgp,
Hazra:2022rdl,Antony:2022ert,Silverstein:2008sg,McAllister:2008hb,Behbahani:2011it,Wang:2002hf,Easther:2002xe,Bozza:2003pr,Dvorkin:2010dn,Dvorkin:2009ne,Mortonson:2010er,Adshead:2011bw,Adams:2001vc,Hu:2001fb}
and such oscillations originate from string motivated models
\cite{McAllister:2008hb,Flauger:2009ab}, or other theoretical
frameworks. The inflationary era was initially proposed to solve
several shortcomings of the classical hot Big Bang theory, such as
the flatness problem, the monopoles problem and the horizon
problem. The Universe emerged as a classical regime, but due to
the fact that inflation is time-wise close to the quantum gravity
regime, imprints of this stringy era might be imprinted on the
inflationary era and its observable quantities. Many effects of
oscillations have been examined in the literature
\cite{Zeng:2018ufm,Domenech:2019cyh,Antony:2021bgp} and even in
the Planck collaboration article \cite{Planck:2018jri}, so it is a
well studied subject. In this work we aim to present a different
perspective of inflationary dynamics, related to the finite
duration of the recombination era. The standard CMB analysis
assumes a direct deterministic mapping between the multipole
$\ell$ probed by the CMB and the primordial wavenumber $k$. Since
the recombination era has a finite duration, this mapping is
intrinsically probabilistic. Thus we elevate the power spectrum of
the primordial perturbations to a probability distribution due to
the finite duration of the recombination. We show that finite
recombination width introduces a Gaussian smoothing scale in $\ln
k$ with $\sigma_{\ln k} \sim \sigma_\eta / D_*$, leading to
exponential suppression of high-frequency primordial features and
a probabilistic mapping from multipoles to inflationary e-folds.
This effect is zero in standard power-law inflationary scenarios,
but it becomes relevant in extended recombination scenarios or
high-precision feature searches combined with oscillating features
of the primordial power spectrum. The result of this work is that
the effect of the primordial oscillations may generate observable
effects in the spectral index inferred by the TT and EE modes. In
fact, the difference $n_s^{TT}-n_s^{EE}$ is non-trivial if the
smoothing effects are taken into account. This result is supported
by a minimal Fisher matrix which we perform and show that the
smoothing scale is a new degree of freedom in the eigensystem of
the Fisher matrix.

\section{Smoothing for Inflationary Modes by Extended Recombination Era}

\subsection{Qualitative Discussion of the Core Idea of Inflationary Smoothing}

As we mentioned in the introduction, the aim of this work is to
highlight a hidden assumption in current CMB inference frameworks.
Specifically, we shall alter the deterministic mapping between
angular multipoles $\ell$ and the primordial wavenumbers $k$. The
CMB is not a sharp snapshot at one conformal time $\eta_{*}$, but
if the recombination lasts for a redshift range $\Delta
z_{rec}=80-100$, then the CMB is an integral over a $\eta$-shell
of finite thickness $\Delta \eta$. This distribution of conformal
times can be propagated in the inflationary modes. This assumption
changes crucially the mapping between the inflationary wavenumbers
$k$ and the CMB multipoles from nearly one-to-one, to a
convolution over a range of horizon-crossing conformal times. Now
let us built up the essential features of our work and gradually
proceed to the main result of our analysis.

In the standard mapping $k \leftrightarrow \ell$, inflation
produces a primordial power spectrum,
\begin{equation}
P_\mathcal{R}(k) = A_s \left( \frac{k}{k_*} \right)^{n_s-1}\, ,
\end{equation}
where $n_s$ is the spectral index of the scalar primordial
curvature perturbations, $k_{*}$ is the CMB pivot scale $k_* =
0.05\, \mathrm{Mpc}^{-1}$, and $A_s$ is the amplitude of the
primordial scalar perturbations. Each mode contributes to the
angular power spectrum via the relation,
\begin{equation}
C_\ell = \int \frac{dk}{k}\, P_\mathcal{R}(k)\,
|\Delta_\ell(k)|^2,
\end{equation}
where $\Delta_\ell(k)$ is the radiation transfer function. In the
standard instantaneous recombination approximation we have.
\begin{equation}
\Delta_\ell(k) \approx S(k,\eta_*)\, j_\ell\bigl[k(\eta_0 -
\eta_*)\bigr],
\end{equation}
with all quantities evaluated at a single last-scattering time
$\eta_*$. This leads to the well-known approximate correspondence
\begin{equation}
k \approx \frac{\ell}{D_*},
\end{equation}
with $D_*=\eta_0 - \eta_*$, so that each multipole $\ell$
effectively probes one characteristic inflationary wavenumber,
thus the $k\leftrightarrow \ell$ correspondence. This is why the
parameters are quoted at a single pivot scale $k_*$.

Now let us get into the core of our idea, and we shall assume a
finite recombination time and quantify this as a convolution in
the conformal time. With a realistic finite recombination
duration, the transfer function generalizes to the following,
\begin{equation}\label{visibility1}
\Delta_\ell(k) = \int d\eta\, g(\eta)\, S(k,\eta)\,
j_\ell\bigl[k(\eta_0 - \eta)\bigr]\, ,
\end{equation}
and in conventional settings we have
$g(\eta)=\delta(\eta-\eta_*)$, thus the instantaneous nature of
the recombination is assumed. In the generalized approach we
introduce with a finite range of conformal times, $g(\eta)$ is not
a Dirac delta function anymore, but a distribution of conformal
times. Each multipole $\ell$ now receives contributions from a
range of conformal times $\eta$, and therefore from a range of
horizon sizes at the last scattering surface which now is not a
slice, a 3-dimensional spacelike hypersurface, but a shell of
thickness $\Delta \eta$ comprised by 3-dimensional spacelike
hypersurfaces. Thus the multipole $\ell$ receives contributions
not from one wavenumber corresponding to one conformal time, but
from a range of wavenumbers,
\begin{equation}
k_{\rm hor}(\eta) \sim a(\eta) H(\eta).
\end{equation}
Inflationary modes exit the horizon ($k = aH$) at a different
e-folding time $N(k)$. Normally this mapping is (nearly) unique:
\begin{equation}
N_* \approx 50{-}60 \quad \longleftrightarrow \quad k_*.
\end{equation}
With a finite recombination thickness $\Delta\eta_{\rm rec}$, each
observed multipole $\ell$ receives power from a band of
e-foldings,
\begin{equation}
N \in [N_1(\ell),\ N_2(\ell)].
\end{equation}
Thus, the pivot scale becomes a distribution of wavenumbers, a
finite range of wavenumbers. Thus instead of a sharp one-to-one
correspondence $k_* \leftrightarrow \ell_*$, each multipole probes
a window of wavenumbers
\begin{equation}
\frac{\Delta k}{k} \sim \frac{\Delta\eta_{\rm rec}}{\eta_*}\, ,
\end{equation}
thus the inferred amplitude therefore becomes an average:
\begin{equation}
A_s(k_*) \quad \to \quad \langle A_s \rangle_W.
\end{equation}
In effect, the standard inference chain $\mathrm{Inflation}\to
P(k) \to C_\ell$, is replace by the correspondence
$\mathrm{Inflation}\to P(k) \to \text{convolution in }k \to
C_\ell$. The likelihood therefore depends on the convolved
spectrum,
\begin{equation}\label{convo1}
P_{eff}(k') = \int d\ln k\, P(k)\, W(k,k')\, ,
\end{equation}
where $W(k,k')$ will be found explicitly in the following. This
effect introduces new degeneracies, the running of the spectral
index $n_s$ is related to the recombination duration, the
amplitude and shape of primordial features are related to the
visibility width, and finally, the inferred number of e-folds
$N_*$ becomes probabilistic, of the form $P(N|\ell)$. Hence we
have a new correspondence between inflation physics and the
recombination duration. In effect, the inflationary parameters are
effectively smoothed over a range of scales and e-foldings. Hence,
the CMB ceases to be a precise local measure of the primordial
power spectrum $P(k)$ and it becomes a low-pass filtered, and
smeared version of the primordial spectrum. This is the core idea
of this article, and as we now evince, this effect may have
quantifiable impact on the observational indices if the primordial
spectrum has oscillating behavior. We can quantify these
considerations by explicitly deriving the $k$-space convolution
kernel $W(k,k')$ in Eq. (\ref{convo1}), from some realistic
$g(\eta)$ and we shall calculate $\Delta k / k$ as a function of
$\Delta z_{\rm rec}$ directly.

Before getting to this, let us discuss in brief what is already
known in the literature about the effects of having a finite
duration for the recombination epoch and explain quantitatively
why the approach we shall adopt is different from known phenomena.
It is known that recombination is not instantaneous, thus we
observe photons from a finite-width last scattering surface, which
is standard in cosmology. Specifically, the duration of the
extended recombination is known, $\approx 115{,}000$ years which
corresponds to a comoving thickness $\approx 19$ Mpc, which in
turn, corresponds to $\Delta z \approx 80{-}100$. Hence it is
known in the literature that the CMB is emitted over a range of
conformal times, not from a single hypersurface corresponding to
one conformal time. This finite thickness already causes damping
of small-scale anisotropies and also causes a smoothing of
acoustic peaks and finally it may have polarization signatures.

In addition, the finite last scattering surface thickness combined
with photon diffusion, causes an exponential damping tail. Thus
small-scale modes get averaged because photons random-walk before
last scattering. However, in the literature, this frequency
averaging effect is treated as a plasma microphysics damping
effect and not as an inflationary scale-mapping problem. When
experiments fit the CMB multipoles they assume,
\begin{equation}
C_\ell = \int dk\, P(k)\, |\Delta_\ell(k)|^2
\end{equation}
with a fixed transfer function. In our case, the transfer function
changes due to recombination width changes, and the experimental
fit compensates by changing $P(k)$. This is the key conceptual
point of this article. In ordinary cosmology contexts, the
thickness of the last scattering surface is treated as a small
smoothing kernel, but not as a conceptual uncertainty on the
mapping of the mapping $\ell \leftrightarrow k \leftrightarrow
\eta_*$. The core concept in our approach essentially breaks the
one-to-one correspondence $\ell \leftrightarrow k \leftrightarrow
\eta_*$, and we will essentially treat the mapping $\ell
\leftrightarrow k$ not as one-to-one but rather as a probabilistic
mapping $P(k|\ell)\leftrightarrow P(N|\ell)$. So we will propagate
the visibility function distribution $g(\eta)$ in Eq.
(\ref{visibility1}) into the pivot scale definition. In effect,
the recombination thickness will induce  an uncertainty in the
inflationary observables. Therefore, $W(k,k')$ in Eq.
(\ref{convo1}) will not be treated simply as a transfer function,
but will be treated as an uncertainty in inflation mapping. In
effect, we will not smooth out the anisotropies, but we will
perform a smoothing of the inflationary time inference. This is an
entirely new perspective in inflationary cosmology, in which the
$\ell \leftrightarrow k$ and $N \leftrightarrow k$ one-to-one
mappings to probability distributions.

\subsection{From a Finite Range Recombination Width to a $P(k|\ell)$ Probability Kernel}

Our aim is to turn the finite range visibility function $g(\eta)$
into a kernel $P(k|\ell)$ and propagate the probabilistic
uncertainty into inflationary observables. The key result of our
work is that CMB does not actually measure or constrain the power
spectrum itself $P(k)$, but instead it measures and constrains an
averaged version of it in a range of wavenumbers $\langle
P(k)\rangle_{W_{\ell}(k,k')}$. We start from the multipole
$C_{\ell}$, defined as,
\begin{equation}\label{multipole1}
C_\ell = \int \frac{dk'}{k'}\, P(k')\, |\Delta_\ell(k')|^2,
\end{equation}
with the transfer function $\Delta_\ell(k')$ being,
\begin{equation}\label{visibility2}
|\Delta_\ell(k')|^2 = \int_0^{\eta_0} d\eta\, g(\eta)\,
|S(k',\eta)|^2\, |j_\ell\bigl[k'(\eta_0 - \eta)\bigr]|^2\, ,
\end{equation}
and we define $r=\eta_0-\eta$ and $r_*=\eta_0-\eta_*$, and let
$\delta \eta=\eta-\eta_*$, hence $r=r_*-\delta \eta$, therefore,
$j_{\ell}(k'r)=j_{\ell}(k'(\eta_0-\eta))$. Also consider the
change of variable from $\eta$ to $\chi=\eta_0-\eta$ and the
central relation $k'\chi_*=\ell$. Now since we assumed a finite
duration for the recombination, the actual wavenumber related to
the finite spread of the variable $\eta$ is not $k'$, but the
variable $k$, defined as follows,
\begin{equation}\label{endiamesi}
k=\frac{\ell}{\chi_*+\delta \chi}\, .
\end{equation}
so for small fluctuations $\chi=\chi_*+\delta \chi$ and we expand,
\begin{equation}\label{exp1}
k=\frac{\ell}{\chi_*}\left(1-\frac{\delta\chi}{\chi_*} \right)\, ,
\end{equation}
so $\frac{\delta k}{k}=-\frac{\delta \chi}{\chi_*}$, or
equivalently,
\begin{equation}\label{logfirst}
\delta \ln k=\frac{\delta k}{k}=-\frac{\delta \chi}{\chi_*}\, .
\end{equation}
Thus, $k'=\frac{k}{1+\delta \eta/r_*}$ or by Taylor expanding we
get, $k'=k-k\frac{\delta \eta}{r_*}$, thus we have,
\begin{equation}\label{visibility2extrahamos}
|\Delta_\ell(k')|^2 = \int_0^{\eta_0} d\eta\, g(\eta)\,
|S(k-k\frac{\delta \eta}{r_*},\eta_*+\delta \eta)|^2\,
|j_\ell\bigl[(k-k\frac{\delta \eta}{r_*})(r_*-\delta
\eta)\bigr]|^2\, .
\end{equation}
We shall take a Gaussian visibility function
\begin{equation}\label{gaussianvisibility}
g(\eta)=\frac{1}{\sqrt{2\pi}\sigma_{\eta}}e^{-\frac{(\eta-\eta_*)^2}{2\sigma_{\eta}^2}}\,
,
\end{equation}
where $\sigma_{\eta}\ll 1$ indicates the width of the
recombination finite duration, so with the change of variable we
performed earlier and the transformation of the $\eta$ integration
to a $k$ integration, we have $\frac{\delta k}{k}=-\frac{\delta
\chi}{\chi_*}$, or equivalently,
\begin{equation}\label{logfirst}
\delta \ln k=\frac{\delta k}{k}=-\frac{\delta \chi}{\chi_*}\, .
\end{equation}
Now insert this in the multipole function (\ref{multipole1}), and
we have,
\begin{equation}\label{multipoleprefinal}
C_\ell = \int d\ln k\int d\ln k'\, P(k')|S(k,\eta_*)|^2\,
|j_\ell\bigl[k(\eta_0 - \eta_*)\bigr]|^2
 \,\frac{1}{\sqrt{2\pi}\sigma_{\ln
k}}e^{-\frac{(\ln k-\ln k')^2}{2\sigma_{\ln k}^2}}\, ,
\end{equation}
thus we can recast this as follows,
\begin{equation}\label{multipolefinal}
C_\ell = \int d\ln k P_{eff}(k)|S(k,\eta_*)|^2\,
|j_\ell\bigl[k(\eta_0 - \eta_*)\bigr]|^2\, ,
\end{equation}
where $P_{eff}(k)$ is,
\begin{equation}\label{peff}
P_{eff}(k)=\int d \ln k' P(k')W(k,k')\, ,
\end{equation}
and $W(k,k')$ is defined as,
\begin{equation}\label{kerneldefinition}
W(k,k')=\frac{1}{\sqrt{2\pi}\sigma_{\ln k}}e^{-\frac{(\ln k-\ln
k')^2}{2\sigma_{\ln k}^2}}\, ,
\end{equation}
with normalization,
\begin{equation}\label{normalizationconditionw}
\int \ln k W(k,k')=1
\end{equation}
Note that we defined $\sigma_{\ln
k}=\frac{\sigma_{\chi}}{\chi_*}$. Also note that the resulting
expression in Eq. (\ref{multipolefinal}) is a leading order result
which we obtained after Taylor expanding the expressions
(\ref{visibility2extrahamos}), in terms of $\delta \eta$. Note
also that the result holds true for intermediate and high
multipoles.

Equation (\ref{peff}) is central in our analysis and was obtained
by propagating the $\eta$ uncertainty to the $k'$ uncertainty. The
is another more formal way to obtain Eq. (\ref{peff}) which we now
quote. The primordial perturbations are a random field,
\begin{equation}
\langle \zeta(\mathbf{k})\,\zeta(\mathbf{k}')\rangle = (2\pi)^3
\delta^{(3)}(\mathbf{k} + \mathbf{k}')\, P(k)
\end{equation}
So $P(k)$ is basically the variance of the Fourier modes, a
stochastic process and lives naturally in logarithmic wavenumber
space.

Therefore $P(k)$ is not an observable directly, it is a latent
spectrum. Any cosmological measurement reconstructs a scale from
angular data. We define, $X = \ln k$ which is the true
inflationary variable, and $Y = \ln k'$ the reconstructed scale
inferred from angular multipoles. Due to projection effects and
finite radial thickness, the mapping between true $k$ and
reconstructed $k'$ is not deterministic and thus we introduce a
random variable,
\begin{equation}
Y = X + \epsilon
\end{equation}
where $\epsilon$ is a random variable with mean zero and finite
variance $\sigma^2$. We assume,
\begin{equation}
\epsilon \sim f_\epsilon(\epsilon) \qquad\text{with}\qquad \int
d\epsilon \, f_\epsilon(\epsilon) = 1.
\end{equation}
The conditional probability is then,
\begin{equation}
p(Y \mid X) = f_\epsilon(Y - X).
\end{equation}
The effective spectrum inferred at the reconstructed scale $Y$ is
the expectation value of the true power at all possible true
scales that could produce this reconstructed scale,
\begin{equation}
P_{\mathrm{eff}}(Y) = \mathbb{E}\bigl[ P(X) \,\big|\, Y \bigr].
\end{equation}
Using marginalization,
\begin{equation}
P_{\mathrm{eff}}(Y) = \int dX \, P(X) \, p(Y \mid X) = \int dX \,
P(X) \, f_\epsilon(Y - X),
\end{equation}
and by turning to $k$ variables, we get,
\begin{equation}
P_{\mathrm{eff}}(k) = \int d\ln k' \, P(k') \, W(k, k'),
\end{equation}
where the convolution kernel is,
\begin{equation}
W(k, k') = f_\epsilon(\ln k - \ln k'),
\end{equation}
which is a pure probability convolution. In this case, we did not
invoke the conformal time, but we only assumed that the mapping
between the true scale and the inferred scale contains additive
noise in logarithmic space. The question is why additive? The
angular projection relates scales multiplicatively:
\begin{equation}
\ell \sim k r\, ,
\end{equation}
and if the effective distance $r$ fluctuates slightly, $k \to k(1
+ \delta)$, then taking logarithms, we get,
\begin{equation}
\ln k \;\to\; \ln k + \delta.
\end{equation}
Thus the noise is therefore additive in $\ln k$. Also we choose a
Gaussian kernel, for the reason that, the uncertainty in the
reconstructed scale arises from many small independent effects, a
finite radial thickness, the projection geometry, the transfer
function width. Thus the sum of many small fluctuations is a
Gaussian. Plus, it is the simplest choice for random variables.
Thus the natural choice is,
\begin{equation}
W(k,k') = \frac{1}{\sqrt{2\pi}\sigma_{\ln k}} \exp\left[
-\frac{(\ln k - \ln k')^2}{2\sigma_{\ln k}^2} \right].
\end{equation}
Therefore, the observed effective spectrum is the true primordial
spectrum blurred by uncertainty in scale reconstruction which is
mathematically identical to Bayesian marginalization over a latent
variable and propagation of the measurement error in the
independent variable. This behavior is unavoidable if the
independent variable ($k$ or $\ln k$) is itself noisy.

Let us proceed to a quantitative level and we shall try to make
contact of the effective primordial perturbation with the
inflationary phenomenological observables. To this end, we recall
the effective primordial spectrum,
\begin{equation}\label{effprimodrdialexplicit}
P_{eff}(k)=\int_0^{\infty}d \ln k' P(k')W(k,k')\, ,
\end{equation}
and we change the variable to $x=\ln k$ and $x'=\ln k'$, thus we
get,
\begin{equation}\label{effprimodrdialexplicit}
P_{eff}(x)=\int_{-\infty}^{\infty}dx P(x')W(x,x')\, ,
\end{equation}
with $\int_{-\infty}^{\infty}dx W(x,x')=1$, and recall also that
$W(x,x')$ is sharply peaked around $x'-x=0$, with a width
$\sigma_x\ll 1$. We define $y=x'-x$ so we have,
\begin{equation}\label{epeffnew}
P_{eff}(x)=\int_{-\infty}^{\infty}W(y)P(x+y)d y\, ,
\end{equation}
since $y$ is narrowly peaked, we Taylor expand, and we get,
\begin{equation}\label{taylorexp}
P(x+y)=P(x)+y\frac{dP}{dx}\Big{|}_{y=0}+\frac{y^2}{2}\frac{d^2P}{dx^2}\Big{|}_{y=0}+...\,
,
\end{equation}
and we insert the expression (\ref{taylorexp}) in the integral
(\ref{epeffnew}), and therefore we have,
\begin{equation}\label{prefinalexprpeff}
P_{eff}(x)=\int dy
W(y)\left(P(x)+y\frac{dP}{dx}\Big{|}_{y=0}+\frac{y^2}{2}\frac{d^2P}{dx^2}\Big{|}_{y=0}+...
\right)\, ,
\end{equation}
so we have,
\begin{align}\label{resultsofintegration}
& \int_{-\infty}^{\infty} dy W(y)=1\, , \\ \notag &
\int_{-\infty}^{\infty} dy y\,W(y)=0\, , \\ \notag &
\int_{-\infty}^{\infty} dy y^2 W(y)=\sigma_x^2\, ,
\end{align}
thus we obtain at leading order,
\begin{equation}\label{prefinalexprpeff1}
P_{eff}(x)=P(x)\left(1+\frac{\sigma_x^2}{2}\frac{P''}{P}\right)\,
,
\end{equation}
where the prime denotes differentiation with respect to $x$. Upon
taking the logarithm, we have,
\begin{equation}\label{logtricks1}
\ln P_{eff}(x)=\ln
P(x)+\ln\left(1+\frac{\sigma_x^2}{2}\frac{P''}{P}\right)\, ,
\end{equation}
so by expanding for small $\sigma_{x}$, we have $\ln
(1+\epsilon)=\epsilon-\frac{\epsilon^2}{2}$, with
$\epsilon=\frac{\sigma_x^2}{2}\frac{P''}{P}$, so we have,
\begin{equation}\label{isp}
\ln P_{eff}(x)=\ln P(x)+\frac{\sigma_x^2}{2}\frac{P''}{P}\, .
\end{equation}
Now using the identity,
\begin{equation}\label{identity}
\frac{P''}{P}=\left(\ln P \right)''+\left((\ln P)'\right)^2\, ,
\end{equation}
and also by assuming a slowly varying power spectrum with
$\frac{P'}{P}\ll 1$, we finally get,
\begin{equation}\label{masterequationprefinal}
\ln P_{eff}(x)=\ln P(x)+\frac{\sigma_x^2}{2}\frac{d^2 \ln
P}{dx^2}\, ,
\end{equation}
which can be rewritten,
\begin{equation}\label{masterequationfinal}
\ln P_{eff}(k)=\ln P(k)+\frac{\sigma_{\ln k}^2}{2}\frac{d^2 \ln
P}{d(\ln k)^2}\, .
\end{equation}
The equation (\ref{masterequationprefinal}) is the major result of
this article and we shall utilize this in order to find any
phenomenological implications for inflation. The result basically
changes the perspective in inflationary cosmology through the
prism of an effective smoothed primordial power spectrum, standard
inflation assumes the correlation
\begin{equation}
C_\ell \leftrightarrow P(k_\ell)\, ,
\end{equation}
and according to this work, the multipole is correlated to an
averaged power spectrum,
\begin{equation}
C_\ell \leftrightarrow \langle P(k) \rangle_{P(k \mid \ell)}.
\end{equation}
This introduces a new fundamental smoothing scale in $\ln k$,
which propagates into inflationary observables parameter
inference. Let us discuss in brief where we expect to find
observational hints. Firstly, it is obvious that a pure power-law
spectrum cannot yield any observable effect, since the higher
derivatives $\frac{\sigma_{\ln k}^2}{2}\frac{d^n \ln P}{d(\ln
k)^n}$ are simply zero for a pure power-law spectrum. The only
possibility to find observable effects is if the spectrum contains
oscillatory features of the form,
\begin{equation}\label{powerlawoscillations}
P(k)=P_0(k)\left(1+A \cos\left(\omega \ln \frac{k}{k_*}+\phi
\right)\right)\, ,
\end{equation}
with $P_0(k)=A_s\left(\frac{k}{k_*} \right)^{n_s-1}$. Hence we
focus in the case of the oscillatory power spectrum of Eq.
(\ref{powerlawoscillations}), which as we mentioned in the
introduction can be produced by various string motivated
theoretical frameworks. So what do we expect theoretically for
this sort of spectrum? Planck fits $\Lambda$CDM, but it has no
parameter describing the smoothing we introduced in this work. We
expect the distortion of the smoothing to affect the scalar
spectral index tilt $n_s$, the running $\alpha_s$, and more
importantly, some measurable difference in $n_s^{EE}-n_s^{TT}$.
The signal might already be there, but it remains undetectable in
Planck. On the other hand, it might be hard to detect such effects
because Planck is noise-limited at high multipoles $\ell$. The
smoothing effect grows toward small angular scales, so we expect
it to affect multipoles with $\sim \ell^2$, for $\sigma_{\ln k}
\sim 10^{-3}$, hence it mainly affects $\ell \gtrsim 1500$, in
which case the Planck precision is limited.

Now before closing, let us discuss what happens with the
superhorizon modes in the context of our approach. The freezing of
these modes at first horizon crossing is unquestionable, but when
these enter the horizon they start oscillating. Specifically, when
they reenter the horizon, modes start oscillating:
\begin{equation}
\delta_\gamma(k,\eta) \sim \cos(k c_s \eta)\, ,
\end{equation}
and this produces acoustic oscillations. The observed temperature
anisotropy depends on the phase of these oscillations at the last
scattering surface,
\begin{equation}
\Theta_\ell(k) \sim \cos(k r_s(\eta_*)).
\end{equation}
Hence, the observables depend on the time of the last scattering.
If the recombination was instantaneous, we would observe each mode
at a single phase of oscillation. However, in reality
recombination lasts $\Delta t$. Thus we observe each mode at many
phases,
\begin{equation}
\cos(k c_s \eta) \quad \text{with } \eta \in [\eta_* - \Delta\eta,
\eta_* + \Delta\eta]\, ,
\end{equation}
so averaging over time we get,
\begin{equation}
\langle \cos(k c_s \eta) \rangle = \cos(k c_s \eta_*) \, e^{-k^2
c_s^2 \sigma_\eta^2 / 2}\, ,
\end{equation}
which is the acoustic damping factor. Hence, the perturbation
itself does not evolve differently, but the time at which we
observe is uncertain. Since, inflation inference relies on the
chain, \emph{Inflation time} $\to k \to \ell$, the extended
recombination breaks the last step, thus the correlation $k \to
\ell$ becomes probabilistic. Hence, the inflation time of origin
of a multipole becomes probabilistic. This blurs somewhat the
mapping from the angular scale to the primordial scale $k$.

\subsection{Finite Width of the Recombination Last Scattering Surface, Previous Analysis and the New Concept of Inflationary Smoothing}

Let us further elaborate on the core idea of this article, and
discuss the differences between inflationary smoothing and the
finite range recombination width already taken into account in the
literature and used in Boltzmann codes like CMB-Fast and its many
successors, such as CLASS. A central point of the potential
confusion concerns the distinction between what the Boltzmann
solvers compute and how their outputs can be interpreted in the
parameter inference. Let us try to make clear the new facts that
inflationary smoothing brings along and how this is distinguished
from the already known finite width of the recombination era
effects already used in Boltzmann codes. So firstly let us explain
more clearly what the Boltzmann codes calculate. Modern Boltzmann
codes like CLASS and CAMB evaluate the angular power spectrum via,
\begin{equation}
C_\ell^{\mathrm{th}} = \int d\ln k \; P(k)\, |\Delta_\ell(k)|^2,
\end{equation}
with $P(k)$ being the primordial power spectrum, which serves as
input in the code, and $\Delta_\ell(k)$ is the photon transfer
function. Now the transfer function is given by the line-of-sight
solution,
\begin{equation}
\Delta_\ell(k) = \int d\eta \, S(k,\eta)\, j_\ell\!\left[k(\eta_0
- \eta)\right],
\end{equation}
with $S(k,\eta)$ being the source function and $j_\ell$ the
spherical Bessel function.  The source contains the visibility
function,
\begin{equation}
g(\eta) = \dot{\tau} e^{-\tau},
\end{equation}
which of course has a finite width, of the order $\Delta z \sim
80$ or $\sigma_\chi \sim 15\text{--}20~\mathrm{Mpc}$. In effect,
the kernel,
\begin{equation}
K_\ell(k) \equiv |\Delta_\ell(k)|^2
\end{equation}
is explicitly broad in $\ln k$, and more importantly the mapping,
\begin{equation}
P(k) \;\longrightarrow\; C_\ell
\end{equation}
is basically an integral transform of the following form,
\begin{equation}
C_\ell = \int d\ln k \; P(k)\, K_\ell(k).
\end{equation}
Notably, at no stage does the Boltzmann solver assigns a unique
wavenumber $k$ to a given multipole $\ell$. Now let us proceed by
discussing how likelihood analyses do parameter inferences and how
the implicit identification occurs. Specifically, likelihood
analyses like Planck, one fits parameters of the theory like
$\theta = \{A_s, n_s, \alpha_s, \dots\}$ as follows,
\begin{equation}
\mathcal{L}(\theta) \propto \exp\left[ -\frac{1}{2}
(C_\ell^{\mathrm{obs}} - C_\ell^{\mathrm{th}}(\theta))^T
\mathrm{Cov}^{-1} (C_\ell^{\mathrm{obs}} -
C_\ell^{\mathrm{th}}(\theta)) \right].
\end{equation}
The primordial spectrum is typically parameterized as follows,
\begin{equation}
P(k) = A_s \left(\frac{k}{k_*}\right)^{n_s - 1 +
\frac{1}{2}\alpha_s \ln(k/k_*) + \cdots}.
\end{equation}
The interpretation of constraints then implicitly assumes,
\begin{equation}
k_\ell \approx \frac{\ell}{\chi_*},
\end{equation}
and basically, one explicitly states that multipoles around $\ell$
probe the scale $k_\ell$ and $n_s$ is the slope at scale $k_*$.
This is somewhat an implicit assumption. To further highlight it,
let us seek the origin of the $\ell \leftrightarrow k$
approximation. The identification
\begin{equation}
k_\ell \approx \frac{\ell}{\chi_*}
\end{equation}
basically follows from the peak of the spherical Bessel function
$j_\ell(x)$ which peaks at $ x \sim \ell$, which leads to the
approximation,
\begin{equation}
K_\ell(k) \approx \delta\!\left(k - k_\ell\right),
\end{equation}
and therefore,
\begin{equation}
C_\ell \approx P(k_\ell)\, \int d\ln k\, K_\ell(k).
\end{equation}
This is a conceptual approximation, not what is numerically
implemented. At this point, our smoothing idea comes in, and since
recombination has a finite width, $\eta = \eta_* + \delta\eta$, we
have,
\begin{equation}
k(\eta_0 - \eta) = k(\chi_* + \delta\chi).
\end{equation}
Hence, for a fixed $\ell$, the peak condition becomes,
\begin{equation}
k \sim \frac{\ell}{\chi_* + \delta\chi},
\end{equation}
which implies a spread,
\begin{equation}
\frac{\delta k}{k} \sim \frac{\delta\chi}{\chi_*},
\end{equation}
or numerically,
\begin{equation}
\sigma_{\ln k} \sim \frac{\sigma_\chi}{\chi_*} \sim 10^{-2}.
\end{equation}
Hence, $ K_\ell(k)$  is a broad kernel and  not a delta function.
Thus to better understand the inflationary smoothing, one must use
a correct probabilistic interpretation. The correct statement is
not,
\begin{equation}
C_\ell \leftrightarrow P(k_\ell),
\end{equation}
but the following,
\begin{equation}
C_\ell \leftrightarrow \int d\ln k \; P(k)\, P(k|\ell),
\end{equation}
with
\begin{equation}
P(k|\ell) \sim K_\ell(k).
\end{equation}
Hence, each multipole probes a distribution of wavenumbers and not
a specific wavenumber. This basically introduces an effective
smoothing of $P(k)$. If we define the effective spectrum as in the
previous section,
\begin{equation}
P_{\mathrm{eff}}(k) = \int d\ln k' \, P(k')\, W(k,k'),
\end{equation}
for small $\sigma_{\ln k}$, we showed that a Taylor expansion
yields,
\begin{equation}
\ln P_{\mathrm{eff}}(k) = \ln P(k) + \frac{\sigma_{\ln k}^2}{2}
\frac{d^2 \ln P}{d(\ln k)^2} + \mathcal{O}(\sigma_{\ln k}^4),
\end{equation}
which induces a shift in the inferred scalar tilt,
\begin{equation}
\delta n_s = \frac{\sigma_{\ln k}^2}{2} \frac{d^3 \ln P}{d(\ln
k)^3}.
\end{equation}
This is the difference between the inflationary smoothing and the
spread of the recombination width in the Boltzmann codes. The
Boltzmann codes correctly compute the forward mapping,
\[
P(k) \rightarrow C_\ell,
\]
including the full recombination physics, but the parameter
inference implicitly assumes an inverse mapping,
\[
C_\ell \rightarrow P(k_\ell),
\]
which is based on a one-to-one identification $k_\ell \simeq
\ell/\chi_*$. It is exactly this identification that neglects the
finite width of the projection kernel. Our idea comes along at
this point where we assume that the correct mapping is
probabilistic:
\[
\ell \rightarrow P(k|\ell),
\]
which leads to an intrinsic smoothing of the primordial
inflationary spectrum. This smoothing effectively induces
systematic corrections to the inferred inflationary parameters.
Schematically, we can compactify the above as follows, the
previous approaches assume that
\[
C_\ell = \int d\ln k \; P(k)\, K_\ell(k),
\]
and they infer $C_\ell \leftrightarrow P(k_\ell)$, and our
approach instead assumes that
\[
C_\ell \to \int d\ln k \; P(k)\, P(k|\ell),
\]
hence the standard assumption of the deterministic mapping $k \to
\ell$ is basically replaced by the probabilistic relation
$P(k|\ell)$ which is the central result of this article.

Wrapping up the above, at the level of forward modelling the
Boltzmann codes includes the physics of a finite recombination.
However these codes do not include the inference-level assumption.
Indeed, although the forward calculation includes the visibility
function exactly, the interpretation of the resulting $C_\ell$ in
terms of the primordial modes relies on an implicit approximation,
\begin{equation}
\ell \to k \simeq \frac{\ell}{\chi_*}.
\end{equation}
It is exactly this identification which corresponds to treating
the projection as effectively being a localized delta-function,
\begin{equation}
|\Delta_\ell(k)|^2 \to \delta\!\left(\ln k - \ln k_\ell\right),
\end{equation}
with $k_\ell = \ell/\chi_*$. This approximation is not part of the
Boltzmann solver itself, but is implicitly used in a) the
definition of the pivot scale $k_*$, b) the interpretation of
$n_s$ as $d\ln P/d\ln k$ at a given specific scale, c) in the
reconstruction of $P(k)$ from $C_\ell$, and d) the mapping between
$k$ and the inflationary e-foldings number $N$.  On the contrary,
with this work we introduce an explicit projection kernel.
Specifically, our work makes explicit that the mapping between
$\ell$ and $k$ is not deterministic (like a delta function peak),
but probabilistic. Starting from the line-of-sight solution and by
expanding around the peak of the visibility function, we derived
the effective kernel,
\begin{equation}
P(k \mid \ell) = \frac{1}{\sqrt{2\pi}\,\sigma_{\ln k}}
\exp\!\left[ -\frac{\left(\ln k - \ln
k_\ell\right)^2}{2\sigma_{\ln k}^2} \right],
\end{equation}
with a width,
\begin{equation}
\sigma_{\ln k} \simeq \frac{\sigma_\chi}{\chi_*} \sim 10^{-2}.
\end{equation}
This leads to the following statement, that $C_\ell$
 probes a distribution of  k, and not a single
mode. Hence, the observable spectrum is therefore a convolution,
\begin{equation}
P_{\mathrm{eff}}(k) = \int d\ln k' \, P(k')\, W(k,k'),
\end{equation}
with $W$ being directly related to $P(k|\ell)$ as we showed
earlier. Hence in the previous section we expanded for small
values of $\sigma_{\ln k}$ and we obtained,
\begin{equation}
\ln P_{\mathrm{eff}}(k) = \ln P(k) + \frac{\sigma_{\ln k}^2}{2}
\frac{d^2 \ln P}{d(\ln k)^2} + \mathcal{O}(\sigma_{\ln k}^4),
\end{equation}
which in turn induced a shift in the scalar tilt,
\begin{equation}
\delta n_s = \frac{\sigma_{\ln k}^2}{2} \frac{d^3 \ln P}{d(\ln
k)^3}.
\end{equation}
Now a based question is why this feature is not redundant in the
Boltzmann codes. The key distinctions  between our approach and
what is already computed in Boltzmann codes is summed up in the
following table:
\begin{center}
\begin{tabular}{c|c}\label{tablesum}
\textbf{In Boltzmann codes} & \textbf{Our Approach} \\
\hline
Forward calculation $P(k) \to C_\ell$ & Inference mapping $C_\ell \to P(k)$ \\
Includes $g(\eta)$ exactly & Extracts effective $P(k|\ell)$ \\
No reinterpretation of $k$ & Introduces probabilistic $k$ \\
Numerical integration & Analytic kernel + bias formulas
\end{tabular}
\end{center}
Hence, the inflationary smoothing effect is not a missing physical
ingredient in the Boltzmann solvers, but it is a missing
conceptual step in the interpretation of the observables. In
standard likelihood analyses, one assumes that,
\begin{equation}
C_\ell^{\mathrm{th}} = \int d\ln k\, P(k)\, |\Delta_\ell(k)|^2,
\end{equation}
while we proposed replacing $P(k)$ with a smoothed spectrum,
\begin{equation}
P(k) \;\longrightarrow\; P_{\mathrm{eff}}(k;\sigma_{\ln k}),
\end{equation}
or equivalently by inserting a kernel,
\begin{equation}
C_\ell = \int d\ln k \int d\ln k'\, P(k')\, W(k,k')\,
|\Delta_\ell(k)|^2.
\end{equation}
In effect this introduces a new parameter $\sigma_{\ln k}$, which
encodes the effective projection width. We need to note that the
insightful previous studies of oscillatory features
\cite{Dvorkin:2010dn,Dvorkin:2009ne,Mortonson:2010er,Adshead:2011bw,Chen:2008wn,Adams:2001vc,Hu:2001fb}
correctly included projection effects numerically. However, they
did not isolate the projection as a universal kernel and they did
not derive analytic bias formulas for $n_s$. Also, they did not
propagate this effect into the $k \to N$ mapping. Our work is
therefore complementary, since it provides a model-independent
description, it identifies a systematic effect on parameter
inference and it connects recombination physics directly to
inflationary reconstruction. To be specific on what was done in
the previous literature for oscillating spectrums
\cite{Dvorkin:2010dn,Dvorkin:2009ne,Mortonson:2010er,Adshead:2011bw,Chen:2008wn,Adams:2001vc,Hu:2001fb},
let us discuss this further. The previous literature answered a
different question, compared to our article. Specifically,
previous works on oscillatory spectrums answered the question, if
one gives $P(k)$ with oscillatory features, what is the resulting
$C_\ell$? On the contrary, in this work, we asked a different
question, given that $C_\ell$ is a projection, what does it
actually measure about the power spectrum $P(k)$? So basically
these are two entirely different problems. The literature
addresses the forward problem quantified by the relation $P(k) \to
C_\ell$, while we consider the inverse inference problem $C_\ell
\to  P(k)$. This distinction has direct implications for parameter
inference. The previous analyses effectively compute,
\begin{equation}
C_\ell = \int d\ln k \, P(k)\, K_\ell(k),
\end{equation}
with $K_\ell(k) \equiv |\Delta_\ell(k)|^2$ being the transfer
kernel, and they correctly observe that, rapid oscillations in
$P(k)$ are smoothed in $C_\ell$, and also that the smoothing scale
is set by the projection effects and recombination physics.
However, what is missing in the literature and is discussed in
this work, is that we extract an effective conditional kernel
$P(k|\ell) \propto K_\ell(k)$ and we quantified its width in
log-space  $\sigma_{\ln k} \sim \frac{\sigma_\chi}{\chi_*}$, with
$\sigma_\chi$ being the comoving thickness of the last scattering
surface. Furthermore, we propagated this inference by defining an
effective smoothed spectrum of the form,
\begin{equation}
P_{\rm eff}(k) = \int d\ln k'\, P(k')\, W(k,k'),
\end{equation}
and we derived the induced parameter shifts, for example in the
spectral index,
\begin{equation}
\delta n_s \sim \frac{\sigma_{\ln k}^2}{2} \frac{d^3 \ln P}{d(\ln
k)^3}.
\end{equation}
Hence, this work may considered as a complement of previous works,
since we formulated the projection explicitly as a probabilistic
kernel $P(k|\ell)$, we derived analytic expressions for its width
$\sigma_{\ln k}$, we demonstrated that the standard mapping $\ell
\to k$ is intrinsically probabilistic, we propagated this effect
into biases and uncertainties on inflationary parameter
estimation,  and finally we showed that these biases and
uncertainties induce a systematic floor in precision cosmology
which is not explicitly parameterized in current inference
studies.

In conclusion, the standard assumption $k \to \ell \leftrightarrow
N$ is replaced by,
\begin{equation}
P(k \mid \ell), \quad P(N \mid \ell),
\end{equation}
which implies firstly that inflationary observables are
intrinsically smeared, secondly that there is a fundamental
uncertainty floor from recombination physics and lastly that
precision measurements of $n_s$ must account for this effect.

\subsection{Possible Observable Effects on the CMB}

Let us proceed to the analysis of the possible observable effects
that the primordial smoothing can have on the CMB. We shall
consider the oscillating deformed power spectrum of Eq.
(\ref{powerlawoscillations}) and we shall use the master equation
(\ref{masterequationfinal}) which quantifies the effects of the
primordial smoothing. Note that a pure power-law primordial power
spectrum would receive no contribution from the smoothing. Since
the result of Eq. (\ref{masterequationfinal}) was obtained using
perturbations, we shall confine ourselves in small amplitudes $A$
in the range $A=[0,0.023]$ which is also supported by the
literature \cite{Zeng:2018ufm,Domenech:2019cyh,Antony:2021bgp}. We
shall consider frequencies that make $\frac{P'}{P}\ll 1$ for A in
the range $A=[0,0.023]$. However, we shall also consider small
frequencies, which are motivated by string theoretic arguments,
for example small-frequency oscillations with $\omega < 3$ are
motivated in string-motivated axion monodromy models. These arise
naturally as one limit of the microscopic parameters controlling
the axion period, without invoking any fine-tuning or non-stringy
ingredients. Small $\omega < 3$ is not an exotic limit but a
direct consequence of the compactification-dependent axion period
$f$ being on the larger end of the string-allowed range ($f \sim
0.03$--$0.1\,M_\mathrm{Pl}$).
\begin{figure}
\centering
\includegraphics[width=20pc]{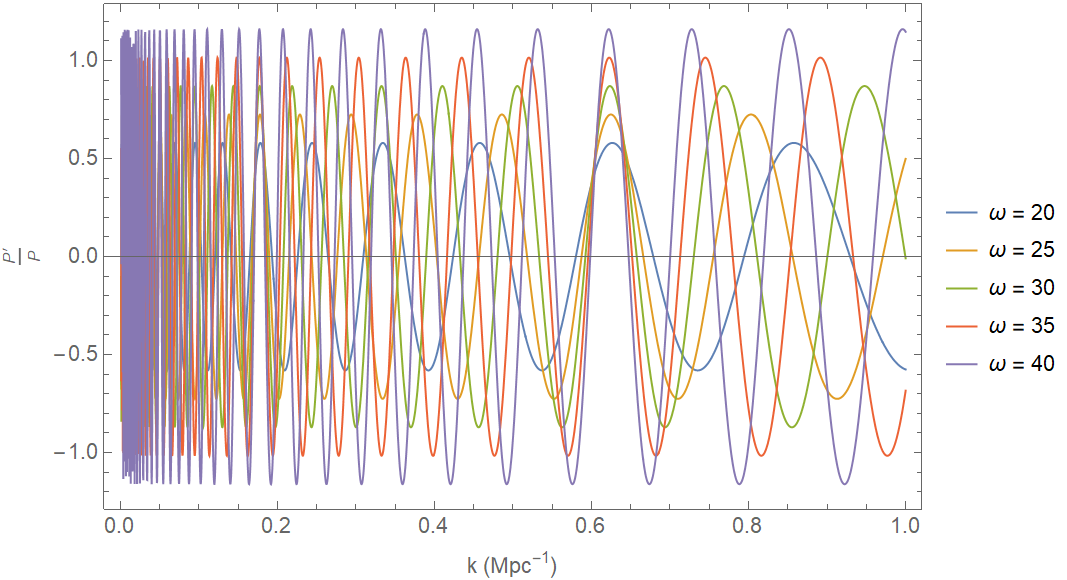}
\includegraphics[width=20pc]{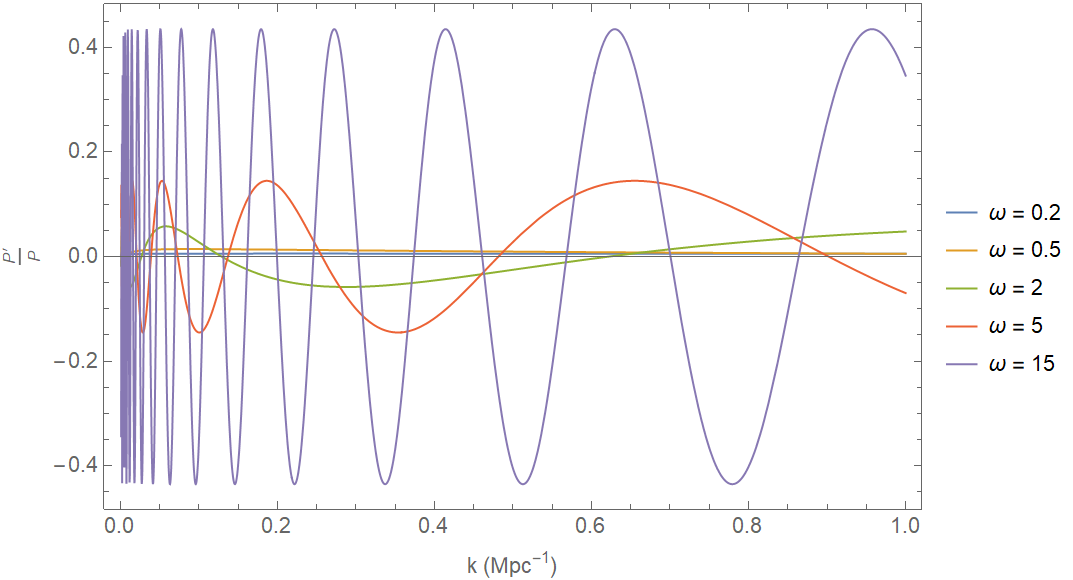}
\includegraphics[width=20pc]{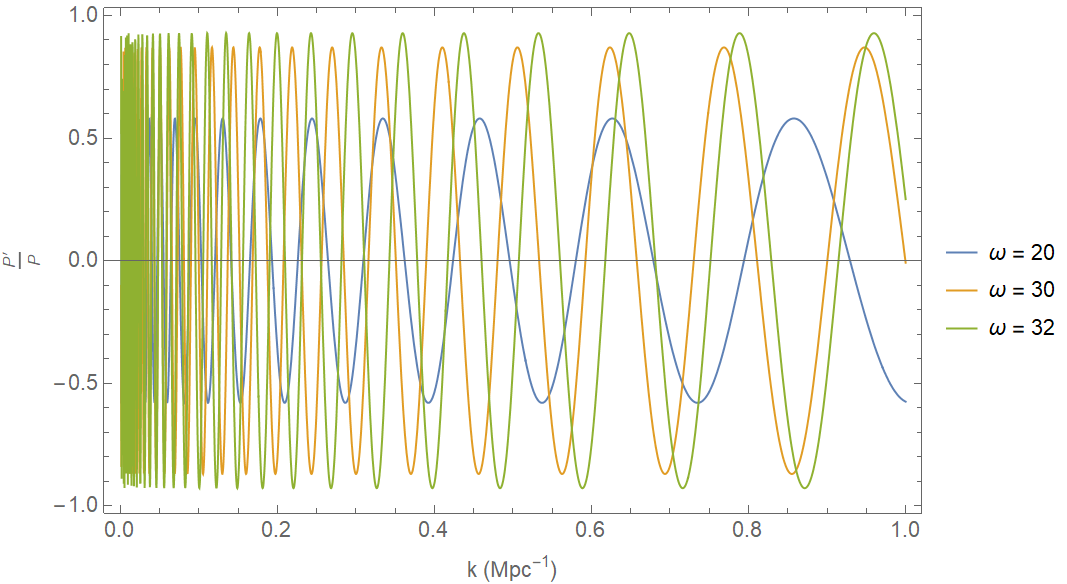}
\caption{The fraction $\frac{P'}{P}$ for $P(k)=P_0(k)\left(1+A
\cos\left(\omega \ln k+\phi-\omega \ln k_* \right)\right)$, and
for $k_*=0.05$Mpc$^{-1}$, $A=0.029$ and $\frac{\phi}{2\pi}=0.698$
in the wavenumber range $k=[0.001,1]$Mpc$^{-1}$, for various
frequencies. The range of frequencies which ensures that
$\frac{P'}{P}\ll 1$ is $\omega=[0,30]$.}\label{plot1}
\end{figure}
It corresponds physically to the inflaton traversing only a small
fraction of one monodromy period during the CMB window, which is
exactly what one expects in a broad class of controlled type IIB
flux/brane constructions. These models remain Planck-viable with
tiny modulation amplitudes and are fully embedded in the same
string framework that produces the more commonly discussed
high-frequency cases \cite{McAllister:2008hb,Flauger:2009ab}.
Future experiments (CMB-S4, LiteBIRD) will be sensitive to this
low-$\omega$ tail. To have a full understanding on the values of
$\omega$ that are allowed, let us plot $\frac{P'}{P}$ as a
function of the wavenumber  for various frequencies in the range
$\omega=[0,40]$. Note that we take $P(k)=P_0(k)\left(1+A
\cos\left(\omega \ln k+\phi-\omega \ln k_* \right)\right)$, with
$k_*=0.05$Mpc$^{-1}$, $A=0.029$ and $\frac{\phi}{2\pi}=0.698$
which are the optimal values for $A$ and $\phi$ from CMB studies
\cite{Zeng:2018ufm}. As it can be seen in Fig. \ref{plot1}, the
optimal frequency range that guarantees $\frac{P'}{P}\ll 1$ is
$\omega=[0,30]$, so we confine ourselves to this range of
frequencies in what follows. Let us proceed to the possible
observable effects in the context of our work, so by using the
master equation (\ref{masterequationfinal}) and the definition of
the spectral index,
\begin{equation}\label{spectralindexdefinition}
n_s^X=1+\frac{d \ln P_X(k)}{d \ln k}\, ,
\end{equation}
the true spectral index is essentially,
\begin{equation}\label{spectralindexdefinition1111}
n_s^{eff}=1+\frac{d \ln P_{eff}(k)}{d \ln k}\, ,
\end{equation}
so in view of Eq. (\ref{masterequationfinal}), the spectral index
becomes,
\begin{equation}\label{spectralindexdefinition222}
n_s^{eff}=1+\frac{d \ln P(k)}{d \ln k}+\frac{\sigma_{\ln
k}^2}{2}\frac{d^3\ln P(k)}{d(\ln k)^3}\, .
\end{equation}
One major prediction of this work, which is absent in the pure
oscillatory spectrum without smoothing, is that there might be
measurable differences in the spectral index of the TT and EE
modes, that is, in the quantity $n_s^{TT}-n_s^{EE}$. Let us
calculate this in detail. We first consider the case that $\omega$
is allowed to take values in the range $\omega=[1,30]$, so
$\omega>1$. For the TT and EE power spectrum for $\omega=[1,30]$
we have,
\begin{align}\label{TTspectrum}
& n_s^{TT}-1=n_s^{PL}-1-A\omega \sin(\omega \ln k-\omega\ln
k_{TT}+\phi_{TT})\\ \notag & +\frac{\sigma_{TT}^{2}}{2}
\Big{[}-\frac{2 A^3 \omega ^3 \sin ^3(\omega \ln k-\omega\ln
k_{TT}+\phi_{TT})}{(A \cos (\omega \ln k-\omega\ln
k_{TT}+\phi_{TT})+1)^3}-\frac{3 A^2 \omega ^3 \sin (\omega \ln
k-\omega\ln k_{TT}+\phi_{TT} ) \cos (\omega \ln k-\omega\ln
k_{TT}+\phi_{TT} )}{(A \cos (\omega \ln k-\omega\ln
k_{TT}+\phi_{TT} )+1)^2}\\ \notag &+\frac{A \omega ^3 \sin (\omega
\ln k-\omega\ln k_{TT}+\phi_{TT} )}{A \cos (\omega \ln k-\omega\ln
k_{TT}+\phi_{TT} )+1}\Big{]}\, ,
\end{align}
\begin{align}\label{EEspectrum}
& n_s^{EE}-1=n_s^{PL}-1-A\omega \sin(\omega \ln k-\omega \ln
k_{EE}+\phi_{EE})\\ \notag & +\frac{\sigma_{EE}^{2}}{2}
\Big{[}-\frac{2 A^3 \omega ^3 \sin ^3(\omega \ln k-\omega\ln
k_{EE}+\phi_{EE})}{(A \cos (\omega \ln k-\omega\ln
k_{EE}+\phi_{EE})+1)^3}-\frac{3 A^2 \omega ^3 \sin (\omega \ln
k-\omega \ln k_{EE}+\phi_{EE} ) \cos (\omega \ln k-\omega \ln
k_{EE}+\phi_{EE} )}{(A \cos (\omega \ln k-\omega\ln
k_{EE}+\phi_{EE} )+1)^2}\\ \notag &+\frac{A \omega ^3 \sin (\omega
\ln k-\omega \ln k_{EE}+\phi_{EE} )}{A \cos (\omega \ln k-\omega
\ln k_{EE}+\phi_{EE} )+1}\Big{]}\, ,
\end{align}
for the spectral indices of the TT and EE spectrums, thus, the
difference $n_s^{TT}-n_s^{EE}$ becomes,
\begin{align}\label{difference}
& n_s^{TT}-n_s^{EE}=\frac{\sigma_{TT}^{2}}{2} \Big{[}-\frac{2 A^3
\omega ^3 \sin ^3(\omega \ln k-\omega \ln k_{TT}+\phi_{TT})}{(A
\cos (\omega \ln k-\omega \ln k_{TT}+\phi_{TT})+1)^3}\\
\notag &-\frac{3 A^2 \omega ^3 \sin (\omega \ln k-\omega \ln
k_{TT}+\phi_{TT} ) \cos (\omega \ln k-\omega \ln k_{TT}+\phi_{TT}
)}{(A \cos (\omega \ln k-\omega \ln k_{TT}+\phi_{TT} )+1)^2}\\
\notag & +\frac{A \omega ^3 \sin (\omega \ln k-\omega \ln
k_{TT}+\phi_{TT} )}{A \cos (\omega \ln k-\omega \ln
k_{TT}+\phi_{TT} )+1}\Big{]}\\\ \notag &-\frac{\sigma_{EE}^{2}}{2}
\Big{[}-\frac{2 A^3 \omega ^3 \sin ^3(\omega \ln k-\omega \ln
k_{EE}+\phi_{EE})}{(A \cos (\omega \ln k-\omega \ln
k_{EE}+\phi_{EE})+1)^3}-\frac{3 A^2 \omega ^3 \sin (\omega \ln
k-\omega \ln k_{EE}+\phi_{EE} ) \cos (\omega \ln k-\omega \ln
k_{EE}+\phi_{EE} )}{(A \cos (\omega \ln k-\omega \ln
k_{EE}+\phi_{EE} )+1)^2}\\ \notag & +\frac{A \omega ^3 \sin
(\omega \ln k-\omega \ln k_{EE}+\phi_{EE} )}{A \cos (\omega \ln
k-\omega \ln k_{EE}+\phi_{EE} )+1}\Big{]}-A\omega \sin(\omega \ln
k-\omega\ln k_{TT}+\phi_{TT})+A\omega \sin(\omega \ln k-\omega\ln
k_{EE}+\phi_{EE})\, ,
\end{align}
where $n_s^{PL}$ is the spectral index corresponding to $P_0(k)$
which is the power-law part of the primordial power spectrum. This
difference between the TT and EE inferred spectral indices
$n_s^{TT}-n_s^{EE}$ is absent in pure power-law power spectrum.
Such an effect is below the sensitivity of current experiments,
but may become detectable with upcoming high-resolution
measurements, such as those from the Simons Observatory or future
CMB Stage-4 experiments. However, such analysis must be performed
with full statistical methods, a task which extends the purposes
of this introductory work. Now we can also obtain a small
frequency limit of the quantity $n_s^{TT}-n_s^{EE}$, so for small
frequencies we have for the EE and TT spectral indices,
\begin{equation}\label{TTspectrum}
n_s^{TT}-1=n_s^{PL}-1-A\omega\left(1-\frac{1}{2}\omega^2\sigma_{TT}^2
\right)\sin(\omega \ln k+\phi_{TT}-\omega \ln k_{TT})\, ,
\end{equation}
\begin{equation}\label{EEspectrum}
n_s^{EE}-1=n_s^{PL}-1-A\omega\left(1-\frac{1}{2}\omega^2\sigma_{EE}^2
\right)\sin(\omega \ln k+\phi_{EE}-\omega \ln k_{EE})\, ,
\end{equation}
From the above two equations (\ref{TTspectrum}) and
(\ref{EEspectrum}), the difference $n_s^{TT}-n_s^{EE}$ is,
\begin{equation}\label{differenceTTee}
n_s^{TT}-n_s^{EE}=-A\omega\left(1-\frac{1}{2}\omega^2\sigma_{TT}^2
\right)\sin(\omega \ln k+\phi_{TT}-\omega \ln
k_{TT})+A\omega\left(1-\frac{1}{2}\omega^2\sigma_{EE}^2
\right)\sin(\omega \ln k+\phi_{EE}-\omega \ln k_{EE})\, .
\end{equation}
The effect of the smoothing on the running of the spectral index
$a_s=\frac{d n_s}{d \ln k}$ must also be taken into account, and
by taking into account the effects of the smoothing, the running
of the spectral index acquires an extra correction term,
\begin{align}\label{runspectralindex}
&\Big{(}\frac{d n_s}{d \ln k}\Big{)}^{smoothed}=\frac{1}{2}
\sigma_{\ln k} ^2 \Big{(}-\frac{3 A^2 \omega ^4 \cos ^2(\omega \ln
k-\omega \ln k_{*}+\phi_{k} )}{(A \cos (\omega \ln k-\omega \ln
k_{*}+\phi_{k})+1)^2}+\frac{4 A^2 \omega ^4 \sin ^2(\omega \ln
k-\omega \ln k_{*}+\phi_{k} )}{(A \cos (\omega \ln k-\omega \ln
k_{*}+\phi_{k} )+1)^2}\\ \notag & -\frac{12 A^3 \omega ^4 \sin
^2(\omega \ln k-\omega \ln k_{*}+\phi_{k} ) \cos (\omega \ln
k-\omega \ln k_{*}+\phi_{k} )}{(A \cos (\omega \ln k-\omega \ln
k_{*}+\phi_{k} )+1)^3}-\frac{6 A^4 \omega ^4 \sin ^4(\omega \ln
k-\omega \ln k_{*}+\phi_{k} )}{(A \cos (\omega \ln k-\omega \ln
k_{*}+\phi_{k} )+1)^4}\\ \notag & +\frac{A \omega ^4 \cos (\omega
\ln k-\omega \ln k_{*}+\phi_{k} )}{A \cos (\omega \ln k-\omega \ln
k_{*}+\phi_{k} )+1}\Big{)}\, .
\end{align}
A comparison between the TT- and EE-derived constraints on the
spectral index $n_s$ may provide a useful consistency test.
Specifically, independent TT-only and EE-only fits may infer
slightly different effective values of the spectral index $n_s$,
because smoothing-induced distortions will project differently
through the temperature and the polarization transfer functions.
The detailed statistical significance of such effects requires a
dedicated analysis, which lies beyond the scope of this
introductory article. We mainly aimed to quantify the effects of a
probabilistic smoothing on the power spectrum caused by an
extended recombination era. A minimal Fisher matrix analysis is
attempted though in the next section, for the general observable
effects of the smoothing scale.

\section{Minimal Fisher Analysis of Log-$k$ Smoothing Effects}

Now in the previous section we showed at an elementary level the
quantitative differences that the primordial modes smoothing
brings along, however we did not discuss the observability
prospects of the smoothing effects in a concrete way. In this
section we shall perform a minimal Fisher analysis in order to
have an elementary grasp of the prospects of observability at a
realistic level. Specifically, we shall focus on the observability
prospects of the primordial smoothing scale $\sigma_{\ln k}$ by
using a minimal Fisher matrix analysis, based on CMB temperature
and polarization anisotropies. Our aim is not to provide a
detailed precision forecast, but to rather to establish whether
the effect could survive the projection into observable
quantities, and in addition to identify the scales at which it
becomes measurable. Our approach is basic, so some further
concrete analysis may be needed to include all the possible
effects that may pinpoint the smoothing effects on the CMB
polarization.

\begin{figure}
\centering
\includegraphics[width=20pc]{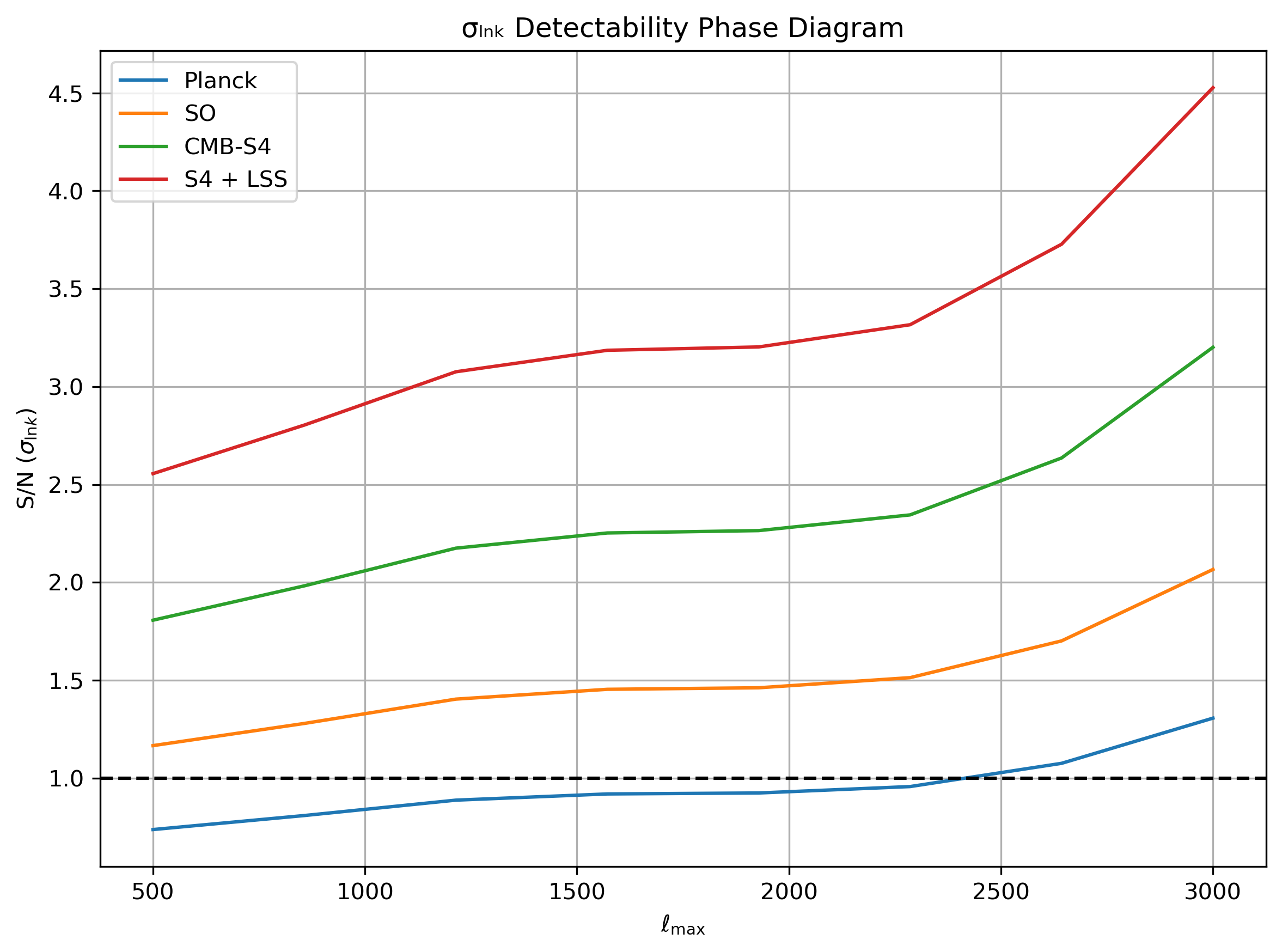}
\caption{The signal-to-noise ratio for the smoothing eigenmode,
for various CMB experiments.}\label{fisherplot}
\end{figure}

As in previous sections, we parameterize the primordial scalar
power spectrum as,
\begin{equation}
\ln P(k) = \ln A_s + (n_s - 1)\ln\frac{k}{k_*} +
\frac{1}{2}\alpha_s \ln^2\frac{k}{k_*} + \frac{1}{6}\beta_s
\ln^3\frac{k}{k_*},
\end{equation}
with the pivot scale  being $k_* = 0.05\,\mathrm{Mpc}^{-1}$. The
smoothing effect is modelled as a Gaussian convolution in the
logarithmic wavenumber,
\begin{equation}
\tilde{P}(k) = \int d\ln k' \, \exp\left[-\frac{(\ln k - \ln
k')^2}{2\sigma_{\ln k}^2}\right] P(k'),
\end{equation}
and equivalently, this can be interpreted as the action of an
operator,
\begin{equation}
\tilde{P}(k) = \exp\left(\frac{1}{2}\sigma_{\ln k}^2 \partial_{\ln
k}^2 \right) P(k),
\end{equation}
which effectively generates an infinite series of higher-order
derivatives in $\ln k$. At a leading order, this induces shifts in
the parameters $(n_s, \alpha_s, \beta_s)$ as we showed in the
previous section, while the subleading terms may produce residual
distortions that cannot be fully absorbed into the finite Taylor
expansion. Now we need to see how these effect project on the CMB
observables. The observable angular power spectra are given by,
\begin{equation}
C_\ell^{XY} = \int \frac{dk}{k} \, P(k)\, \Delta_\ell^X(k)\,
\Delta_\ell^Y(k),
\end{equation}
where $\Delta_\ell^X(k)$ are the radiation transfer functions. Now
even though the smoothing acts simply at the power spectrum $P(k)$
level, the projection kernels differ between the temperature and
polarizations, and they break exact degeneracies. In effect, the
smoothing effect may lead to a characteristic and probably small
distortion pattern in the multipoles. To unveil this, we need a
minimal Fisher matrix analysis. We construct the Fisher matrix for
the following parameter set,
\begin{equation}
\theta = \{A_s, n_s, \alpha_s, \beta_s, \sigma_{\ln k}\},
\end{equation}
by using cosmic-variance-limited CMB spectra,
\begin{equation}
F_{ij} = \sum_\ell \frac{2\ell+1}{2} f_{\rm sky} \left[
\frac{\partial C_\ell^{TT}}{\partial \theta_i} \frac{\partial
C_\ell^{TT}}{\partial \theta_j}\frac{1}{C_\ell^{TT\,2}} +
\frac{\partial C_\ell^{EE}}{\partial \theta_i} \frac{\partial
C_\ell^{EE}}{\partial \theta_j}\frac{1}{C_\ell^{EE\,2}} +
\frac{\partial C_\ell^{TE}}{\partial \theta_i} \frac{\partial
C_\ell^{TE}}{\partial \theta_j} \frac{1}{C_\ell^{TE\,2} +
C_\ell^{TT}C_\ell^{EE}} \right].
\end{equation}
The parameter derivatives are calculated numerically by using a
Boltzmann solver, in which we use the smoothed primordial spectrum
as direct input. Now it is important to identify the $\sigma_{\ln
k}$ mode, and since $\sigma_{\ln k}$ is partially degenerate in
the parameter set $(n_s, \alpha_s, \beta_s)$, we will diagonalize
the Fisher matrix and we identify the least constrained eigenmode
of it,
\begin{equation}
F v_\sigma = \lambda_{\rm min} v_\sigma,
\end{equation}
which basically defines the direction in the parameter space which
is most closely aligned with the smoothing effect. The
signal-to-noise ratio is defined as,
\begin{equation}\label{signaltonoise}
\left(\frac{S}{N}\right)^2 = v_\sigma^T F v_\sigma.
\end{equation}
In order to probe the scales which control the sensitivity, we
shall calculate the Fisher matrix as a function of the maximum
multipole $\ell_{\max}$. The resulting signal-to-noise ratio is
shown in Fig. \ref{fisherplot}.
From Fig. \ref{fisherplot} we can see that there is a monotonic
increase of the sensitivity with $\ell_{\max}$, which indicates
the fact that the smoothing predominantly affects the small-scale
(high-$k$) modes. Also it is evident that a transition from
marginal detectability at $\ell_{\max} \lesssim 2000$ to $S/N
\gtrsim 1$ occurs for the higher-resolution experiments. In
addition, there exists a hierarchical ordering between the CMB
experiments (Planck-like, Simons Observatory, CMB-S4), which
indicates an increasing access to small-scale information. Our
results indicate that the smoothing scale $\sigma_{\ln k}$ behaves
(mainly) as a high-$k$ deformation of the primordial spectrum,
which is only weakly constrained by the large-scale anisotropies.
Interestingly however, the smoothing scale becomes gradually
observable as  the smaller angular scales (large multipoles) are
included. Although our analytic considerations of the previous
section showed that the smoothing effects can be partially
absorbed into the redefinitions of the spectral tilt and running,
the projection into the $C_\ell$ space is not exact. The residual
signal corresponds to a non-polynomial distortion in $\ln k$,
which survives the marginalization over the parameters $(n_s,
\alpha_s, \beta_s)$ and this leads to a finite Fisher sensitivity.
Hence, $\sigma_{\ln k}$ can be interpreted as a smoothing scale in
the primordial fluctuations, instead of a standard slow-roll
parameter. Wrapping the result up, in a nutshell we showed in Fig.
\ref{fisherplot} that the primordial smoothing scale $\sigma_{\ln
k}$ becomes observable at $S/N \gtrsim 1$ for large $\ell_{\max}
\gtrsim 2000$, and can reach $S/N \sim 3$-$5$ for CMB-S4-like
sensitivities, when these are combined with Large Scale Structure
(LSS) information. We emphasize that this conclusion is driven by
the residual, non-degenerate imprint of smoothing on the projected
CMB spectra, and not by simple shifts in spectral tilt or the
running parameters.

But our analysis is a minimal Fisher matrix analysis and it has
limitations. Specifically, we treated the CMB as
cosmic-variance-limited up to $\ell_{\max}$, and we completely
neglected instrumental noise and beam effects. Real CMB
experiments may lose sensitivity at the high multipoles, hence the
curves shown in Fig. \ref{fisherplot} slightly overestimate the
constraining power at large $\ell$. Also we did not include in the
analysis shown in Fig. \ref{fisherplot} the lensing reconstruction
and the LSS information.

Moreover the LSS contribution is approximated by a simple
$k$-space weighting (see next subsection), and not by a realistic
survey model. This should be done concretely in order to obtain a
realistic and far more credible result.  In addition, we did not
include nonlinear evolution and baryon effects, and moreover, the
numerical derivatives and the spline interpolation of the
primordial spectrum may introduce small inaccuracies at high
$\ell$. Also our analysis did not include spectral distortion
observables (for example $\mu$-distortions), which would probe
significantly smaller scales of the order,
\begin{equation}
k \sim 10^2 - 10^4 \ \mathrm{Mpc}^{-1}.
\end{equation}
These are beyond the reach of CMB anisotropies ($k \sim 0.1 -
1\,\mathrm{Mpc}^{-1}$). Therefore, the smoothing parameter
$\sigma_{\ln k}$ primarily affects the high-$k$ modes, and the
present constraints should be regarded as being conservative from
a physical point of view. Therefore, our results should be
interpreted rather as an order-of-magnitude estimation of
detectability, and not an precision forecast. Much concrete
analysis is required in order to obtain an exact precision
forecast. Nevertheless, our minimal Fisher analysis demonstrated
that, despite having analytic suppression at the level of spectral
indices as we showed in the previous section, the smoothing
parameter $\sigma_{\ln k}$ induces a residual and non-degenerate
signature in the CMB anisotropy spectra. This signal effectively
is rendered detectable once sufficient small-scale information is
included, and it provides a characteristic observational pattern
of primordial smoothing effects.

It is important to note that although the analytic differences of
$n_s^{TT}$ and $n_s^{EE}$ are somewhat suppressed , the Fisher
matrix analysis yields observable results. This is because the
Fisher matrix does not probe separate effective tilts for the
temperature and the polarizations, but it probes the full response
of the observable angular spectra to the parameter variations.
Hence, the analytic suppression is only partial, and the residual
effects remain observable once these are projected into $C_\ell$
space.

\subsection{Observability Structure, Principal Component Analysis
and Eigenmode Decomposition}

To further quantify how the smoothing parameter $\sigma_{\ln k}$
affects directly the cosmological observables, we will perform a
principal component analysis and eigenmode decomposition for the
Fisher matrix, now focusing on the eigenmodes. We shall include
contributions from CMB anisotropies, CMB lensing, and simplified
LSS effects. The purpose of this analysis is again not to produce
a precision forecast, but to identify the directions in parameter
space that are actually constrained by the data.

We again consider the parameter set
\begin{equation}
\theta = \{A_s, n_s, \alpha_s, \beta_s, \sigma_{\ln k}\},
\end{equation}
and we construct a total Fisher matrix of the form,
\begin{equation}
F_{\rm tot} = F_{\rm CMB} + F_{\rm lens} + F_{\rm LSS}.
\end{equation}
The CMB contribution is computed from the temperature and the
polarization spectra,
\begin{equation}
F_{ij}^{\rm CMB} = \sum_\ell \frac{2\ell+1}{2} f_{\rm sky} \left[
\frac{\partial C_\ell^{TT}}{\partial \theta_i} \frac{\partial
C_\ell^{TT}}{\partial \theta_j} \frac{1}{C_\ell^{TT\,2}} +
\frac{\partial C_\ell^{EE}}{\partial \theta_i} \frac{\partial
C_\ell^{EE}}{\partial \theta_j} \frac{1}{C_\ell^{EE\,2}} +
\frac{\partial C_\ell^{TE}}{\partial \theta_i} \frac{\partial
C_\ell^{TE}}{\partial \theta_j} \frac{1}{C_\ell^{TE\,2} +
C_\ell^{TT}C_\ell^{EE}} \right].
\end{equation}
The lensing contribution is modelled as
\begin{equation}
F_{ij}^{\rm lens} = \sum_\ell \frac{1}{\sigma_\ell^2}
\frac{\partial C_\ell^{\phi\phi}}{\partial \theta_i}
\frac{\partial C_\ell^{\phi\phi}}{\partial \theta_j}, \qquad
\sigma_\ell^2 = \frac{2 C_\ell^{\phi\phi\,2}}{(2\ell+1)f_{\rm
sky}}.
\end{equation}
The LSS contribution is approximated by a weighted integral over
the linear power spectrum,
\begin{equation}
F_{ij}^{\rm LSS} = \sum_k w(k) \frac{\partial P(k)}{\partial
\theta_i} \frac{\partial P(k)}{\partial \theta_j}, \qquad w(k)
\propto k^2,
\end{equation}
which captures the increasing statistical weight of small-scale
modes. We diagonalize the total Fisher matrix,
\begin{equation}
F_{\rm tot} v_a = \lambda_a v_a,
\end{equation}
and $v_a$ are its orthonormal eigenvectors and $\lambda_a$ are the
corresponding eigenvalues. The corresponding uncertainties are
\begin{equation}
\sigma_a = \lambda_a^{-1/2}.
\end{equation}
Each eigenvector denotes a linear combination of parameters that
is independently constrained by the data. This provides a
basis-independent characterization of the observability. By
diagonalizing the full Fisher matrix, we get the eigenstates
quoted gradually below, which are ordered by increasing error:
\begin{itemize}
\item \textbf{Mode 1}
\begin{equation}
v_1 = (1.000,\; 0.000,\; 0.000,\; 0.000,\; 0.000), \qquad \sigma_1
\simeq 1.40 \times 10^{-12}.
\end{equation}
This mode is completely aligned with $A_s$ and is essentially
perfectly constrained, which indicates the high precision of the
overall amplitude determination from the CMB measurements.
 \item \textbf{Mode 2}
\begin{equation}
v_2 = (0.000,\; 0.554,\; 0.664,\; 0.503,\; 0.002), \qquad \sigma_2
\simeq 9.18 \times 10^{-4}.
\end{equation}
This direction corresponds to a correlated combination of the
parameters $(n_s, \alpha_s, \beta_s)$ with a negligible projection
on $\sigma_{\ln k}$. This mode basically represents the dominant
constrained shape deformation of the primordial spectrum.
 \item
\textbf{Mode 3}
\begin{equation}
v_3 = (0.000,\; -0.832,\; 0.454,\; 0.318,\; 0.004), \qquad
\sigma_3 \simeq 5.81 \times 10^{-3}.
\end{equation}
This mode corresponds to an orthogonal combination of the spectral
tilt and the running parameters, which again are effectively
independent of $\sigma_{\ln k}$.

\item \textbf{Mode 4}
\begin{equation}
v_4 = (0.000,\; 0.017,\; 0.595,\; -0.804,\; -0.006), \qquad
\sigma_4 \simeq 1.37 \times 10^{-2}.
\end{equation}
This direction embodies the higher-order running effects and
exhibits a small but non-zero deviation from $\sigma_{\ln k}$,
which indicates the breaking of partial degeneracy due to the
smoothing.

\item \textbf{Mode 5}
\begin{equation}
v_5 = (0.000,\; 0.003,\; 0.001,\; -0.007,\; 1.000), \qquad
\sigma_5 \simeq 1.98.
\end{equation}
This mode is almost perfectly aligned with $\sigma_{\ln k}$ and it
represents the only direction in the parameter space, where
smoothing is isolated as a distinct degree of freedom. The large
uncertainty confirms that this component is only weakly
constrained by the combined dataset.
\end{itemize}
The above analysis indicates that $\sigma_{\ln k}$ is not fully
absorbed into the parameters $(n_s, \alpha_s, \beta_s)$ once we
include multiple cosmological probes, and it remains statistically
suppressed compared to the spectral parameters. However, the
residual component of the smoothing transformation exists in a
direction that is not aligned with the parameters $(n_s, \alpha_s,
\beta_s)$, and this component is isolated as the least constrained
eigenmode. The fact that this new fifth mode is dominated by the
smoothing scale  $\sigma_{\ln k}$ indicates that the data are
sensitive to a new real degree of freedom. Also the large
uncertainty indicates the partial degeneracy of this new degree of
freedom. This is the important outcome of this work, since the
eigenmode decomposition demonstrated that $\sigma_{\ln k}$
corresponds to a new independent direction in the parameter space.
The observable signal is not a simple shift in the spectral tilt,
but it is a higher-order distortion which survives the projection
on the cosmological observables. Note that this result is obtained
if multiple probes are combined. When these multiple probes are
omitted, the uncertainty spikes extremely.

\section{Conclusions}

Our core assumption for this work is that, although the CMB
analysis assumes a deterministic mapping between multipole $\ell$
and primordial wavenumber $k$, we distorted this mapping due to
the effects of the finite duration of the recombination era, and
thus this mapping becomes intrinsically probabilistic. Thus the
CMB does not measure the primordial spectrum directly, but it
measures a projection of it through the finite thickness of last
scattering surface. This projection thus induces a smoothing in
$k$-space that propagates into all inflationary observables and
sets a new theoretical systematic floor for next-generation CMB
experiments. Another way to see this smoothing effect, is by
thinking that the observed effective power spectrum is the true
primordial spectrum blurred by the uncertainty in scale
reconstruction, which is mathematically identical to a Bayesian
marginalization over a latent variable, and thus there is a
propagation of the measurement error in the independent variable.
The smoothing effect might lead to a non-trivial difference in the
spectral indices inferred from the TT and EE modes, an effect that
may be measurable in future CMB experiments. We performed a
minimal Fisher matrix analysis, in order to see whether the
suppressed effects of the smoothing at the level of spectral
indices, can have observational signatures in the CMB multipoles.
Our eigenmode decomposition demonstrated that the smoothing scale
$\sigma_{\ln k}$ corresponds to a new independent direction in the
parameter space spanned by $\{A_s, n_s, \alpha_s, \beta_s,
\sigma_{\ln k}\}$. We showed that the smoothing scale may cause a
higher-order distortion which survives the projection on the
cosmological observables. However, this result is obtained if
multiple probes are combined and when these multiple probes are
omitted, the uncertainty spikes extremely. Moreover, this was a
minimal Fisher matrix analysis with many simplifications. The
statistical analysis needed for obtaining realistic and robust
observability prospects, stretches by far beyond the scopes of
this introductory article, in which we aimed to point out the
non-trivial effects of the power spectrum smoothing.

\end{document}